\documentclass[conference]{IEEEtran}
\IEEEoverridecommandlockouts
\usepackage{cite}
\usepackage{amsmath,amssymb,amsfonts}
\usepackage{algorithmic}
\usepackage{graphicx}
\usepackage{textcomp}
\usepackage{xcolor}
\usepackage{subfigure}
\usepackage{multirow}
\usepackage{url}
\usepackage{flushend}
\def\BibTeX{{\rm B\kern-.05em{\sc i\kern-.025em b}\kern-.08em
    T\kern-.1667em\lower.7ex\hbox{E}\kern-.125emX}}
\begin{document}

\title{Evaluating the Performance of Over-the-Air \\Time Synchronization for 5G and TSN Integration}

\author{
\IEEEauthorblockN{Haochuan Shi, Adnan Aijaz, and Nan Jiang}
\IEEEauthorblockA{
\text{Bristol Research and Innovation Laboratory, Toshiba Europe Ltd., Bristol, United Kingdom}\\
firstname.lastname@toshiba-bril.com}
}    

\maketitle

\begin{abstract}
The IEEE 802.1 time-sensitive networking (TSN) standards aim at improving the real-time capabilities of standard Ethernet. TSN is widely recognized as the long-term replacement of proprietary technologies for industrial control systems. However, wired connectivity alone is not sufficient to meet the requirements of future industrial systems. The fifth-generation (5G) mobile/cellular technology has been designed with native support for ultra-reliable low-latency communication (uRLLC). 5G is promising to meet the stringent requirements of industrial systems in the wireless domain. Converged operation of 5G and TSN systems is crucial for achieving end-to-end deterministic connectivity in industrial networks. Accurate time synchronization is key to integrated operation of 5G and TSN systems. To this end, this paper evaluates the performance of over-the-air time synchronization mechanism which has been proposed in 3GPP Release 16. We analyze the accuracy of time synchronization through the boundary clock approach in the presence of clock drift and different air-interface timing errors related to reference time indication. We also investigate frequency and scalability aspects of over-the-air time synchronization.   Our performance evaluation reveals the conditions under which 1 \(\mu\)s or below  requirement for TSN time synchronization can be achieved. 
\end{abstract}


\section{Introduction}
Most industrial automation and control systems require real-time connectivity with stringent timeliness and reliability requirements. Such requirements are typically fulfilled through wired technologies like fieldbus systems and industrial Ethernet. However, such technologies are often proprietary and lack interoperability. Time-sensitive Networking (TSN) is a family of standards within the IEEE 802.1 working group \cite{tsn} with an emphasis on enhancing real-time capabilities of standard Ethernet \cite{TSN_standards}. TSN provides guaranteed data delivery with high determinism and bounded low latency. It is widely recognized as the long-term replacement of conventional wired technologies across a range of industrial domains. 

Wireless technologies offer various benefits for industrial communication. Many emerging industrial applications demand real-time connectivity with a higher degree of flexibility and mobility that is not offered by wired technologies like TSN \cite{3gpp_22_804}. Therefore, TSN is expected to co-exist with wireless technologies in future industrial systems. Extending TSN capabilities to Wi-Fi is a natural step as it belongs to the same family of IEEE 802 standards \cite{wireless_TSN_PIEEE}. Recent studies (e.g., \cite{Hfi_OJIES}) have shown that Wi-Fi can provide real-time and deterministic connectivity with modifications to medium access control (MAC) layer. However, Wi-Fi lacks the necessary flexibility for industrial communication. The fifth-generation (5G) mobile/cellular technology has been designed with native support for diverse applications. It has the capability to provide a unified wireless interface for industrial communication. The ultra-reliable low-latency communication (uRLLC) capability of 5G is particularly attractive for providing TSN-like functionality in the wireless domain. 

Integration of 5G and TSN systems is under active investigation within 3GPP \cite{3gpp_23_734}. Converged operation of 5G and TSN systems provides end-to-end deterministic connectivity over hybrid wired and wireless domains. However, such integration also creates a number of challenges \cite{private_5G}. TSN operates on the principle of network-wide time-aware scheduling. Therefore, accurate time synchronization between 5G and TSN systems is crucial for converged operation. The synchronization accuracy must be 1 \(\mu\)s or below to meet jitter requirements of isochronous real-time industrial applications (e.g., motion control systems) \cite{3gpp_22_804}. 

\subsection{Related Work}
Nasrallah \emph{et al.} \cite{TSN_survey} conducted a comprehensive survey of TSN standards and also investigated ultra-low latency mechanisms in 5G. 3GPP has defined the \emph{bridge model} for 5G and TSN integration wherein the 5G system appears as a black box to TSN entities \cite{3gpp_23_734}. The 5G system handles TSN service requests through its internal protocols and procedures. Omri \emph{et al.} \cite{sync_nr}  discussed the synchronization procedure in 5G new radio (NR) and identified that time-frequency synchronization in the presence of wide-range of new frequencies is the main challenge in 5G NR systems. Zhang \emph{et al.} \cite{TAP} proposed a timing method over air interface based on physical layer signals. Mahmood \emph{et al.} \cite{mahmood2018over} discussed the requirements and challenges of over-the-air time synchronization for uRLLC and also investigated some of the key enablers. In another study, the same authors \cite{mahmood2019time} investigated TA-related timing errors for device-level synchronization and discussed averaging of multiple TAs for improving synchronization accuracy. 


\subsection{Contributions and Outline}
While previous studies have investigated different aspects of 5G and TSN integration, the end-to-end performance of over-the-air time synchronization in the presence of potential sources of inaccuracies remains largely unexplored. This paper aims to fill this gap and investigates the performance of reference time indication through over-the-air time synchronization. We develop an analytic framework for accuracy of over-the-air time synchronization in the presence of clock drift and different errors associated with indication of reference time. Based on the analytic framework, we provide results investigating the impact of different errors on overall time synchronization performance, in terms of accuracy for meeting the 1 \(\mu\)s or below requirement. We also investigate scalability and required frequency of over-the-air time synchronization. 

The rest of the paper is organized as follows. Section \ref{prel} covers the necessary preliminaries time synchronization. 
The system model and analytic framework for over-the-air time synchronization is provided in Section \ref{OTA}. Performance evaluation and key results are presented in Section \ref{perf}. Finally, the paper is concluded in Section \ref{CR}. 

\section{Preliminaries} \label{prel}

\subsection{Time Synchronization in TSN}
\begin{figure}
    \centering
    \includegraphics[scale=0.36]{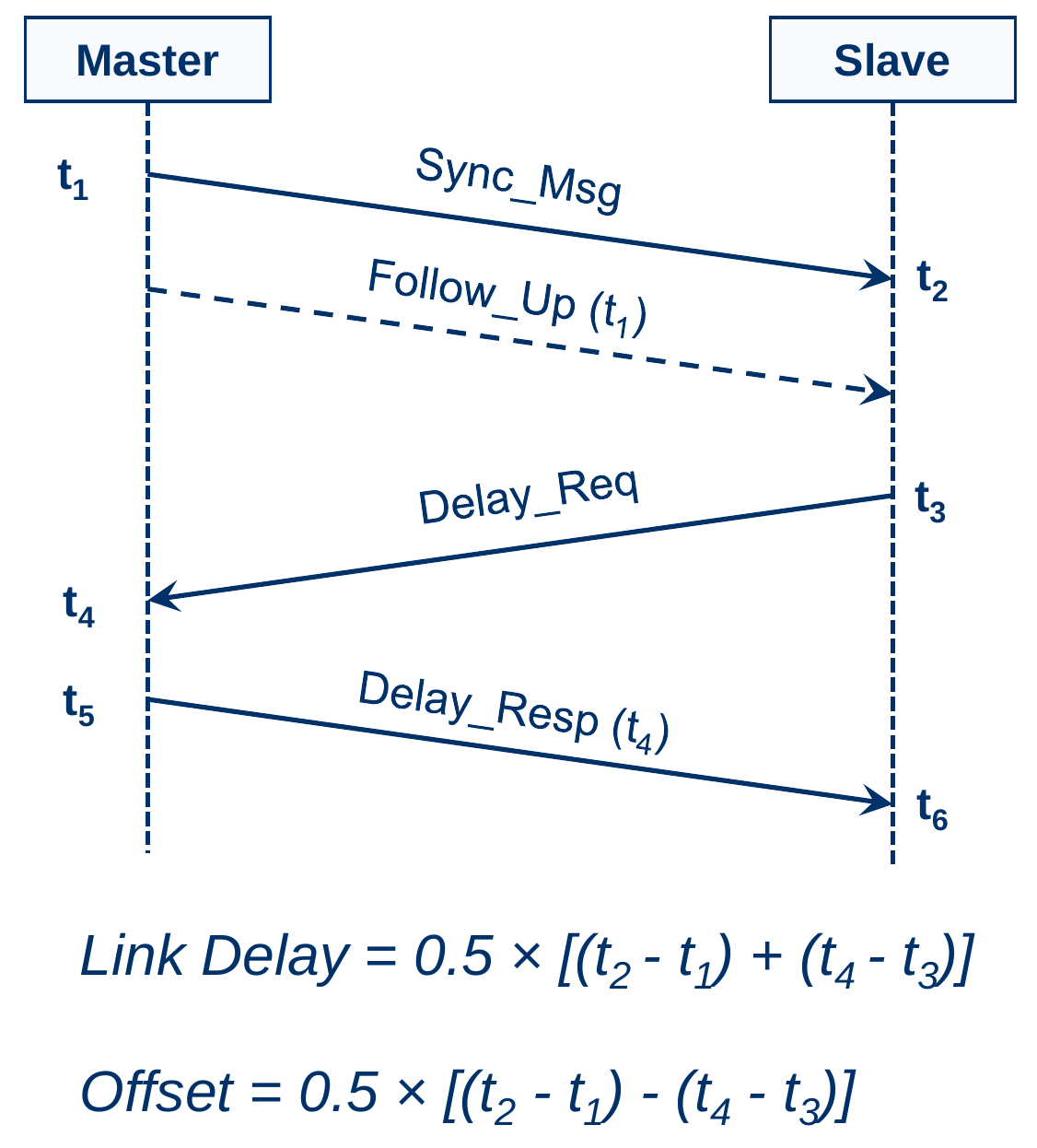}
    \caption{Illustration of PTP operation.}
    \label{ptp}
\end{figure}

Time synchronization in TSN is based on generalized precision time protocol (gPTP) of the IEEE 802.1AS standard \cite{8021as}, which is a profile of the well-known precision time protocol (PTP) defined by the IEEE 1588 standard. PTP defines time synchronization between a master clock and a slave clock. A grandmaster is a master that is synchronized to a time reference. The operation of PTP is illustrated in Fig. \ref{ptp}. Four timestamps are captured between the master clock and the slave clock. These timestamps are used by the slave clock to synchronize to the master clock. Clock difference between the master and the slave is a combination of the clock offset and the message transmission delay. Hence, PTP corrects the clock difference based on offset correction and delay correction. Note that the offset correction assumes that link delay is symmetric, i.e., propagation time from master to slave is equal to the propagation time from slave to master. Since the master and slave clocks drift independently, periodic offset and delay correction are necessary to maintain clock synchronization. The IEEE 802.1AS standard defines a propagation delay measurement procedure between two peer nodes (a peer delay initiator and a peer delay responder) through exchange of timestamps via path delay request and path delay response messages.

\begin{figure}
    \centering
    \includegraphics[scale=0.36]{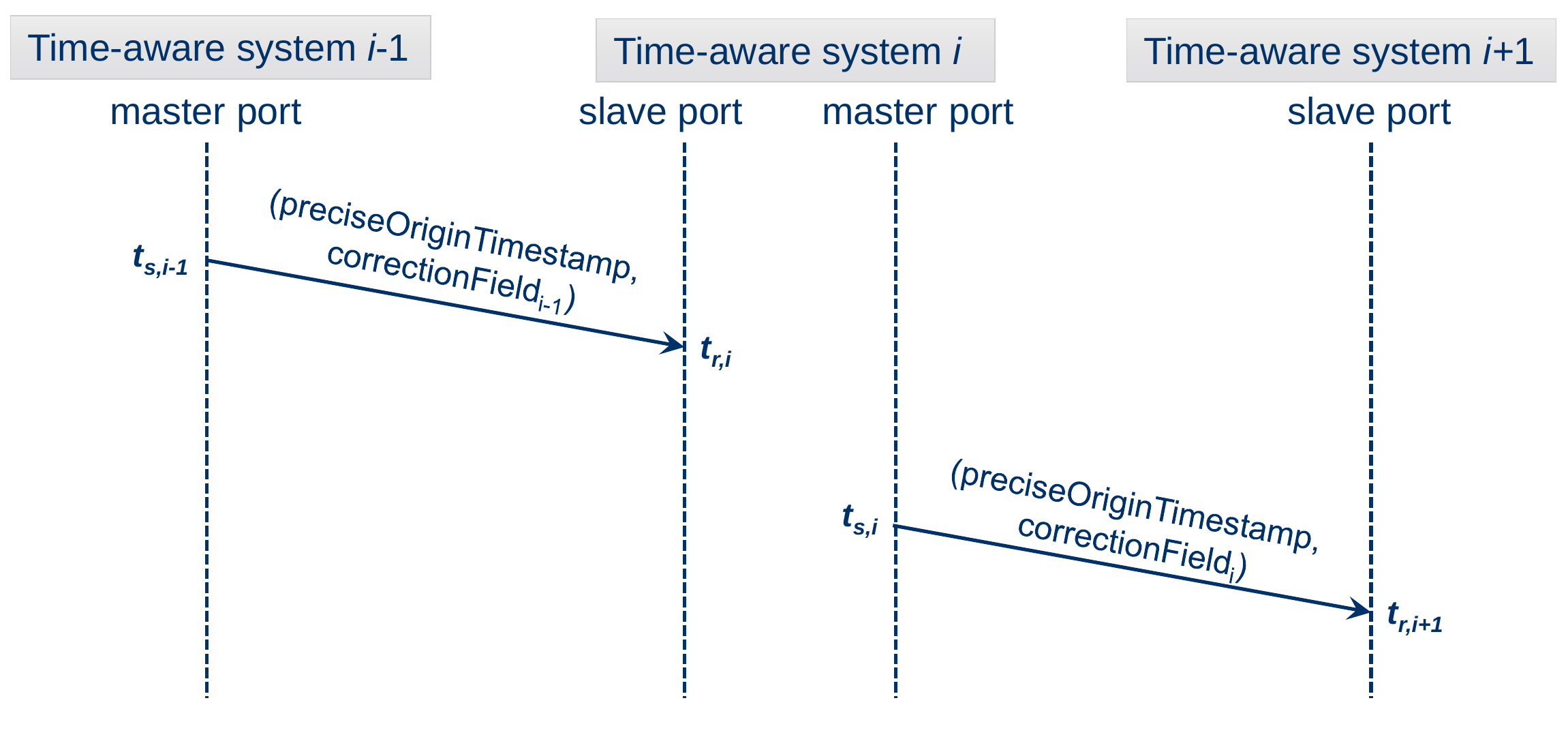}
    \caption{Transport of time synchronization information in IEEE 802.1AS.}
    \label{time_tpt}
\end{figure}

The IEEE 802.1AS standard also specifies transport of time synchronization information over multiple time-aware systems. A TSN bridge receives timing message from its master clock and corrects the error as a slave clock. This bridge then acts as a master clock and forwards the timing information to the next bridge. As illustrated in Fig. \ref{time_tpt}, transport of time synchronization information is achieved through 802.1AS messages containing two key parameters: a preciseOriginTimestamp and  a correctionField. The preciseoriginTimestamp contains the TSN grandmaster time whereas the correctionField provides the residence time in the system. The sum of these two parameters and the link delay  provides synchronized time to the slave.   

\subsection{Clock Models for 5G and TSN Time Synchronization}
For time synchronization in integrated 5G and TSN systems, two main clock models have been considered in 3GPP. These clock models, which are aligned with the IEEE 802.1AS standard, are described as follows. 

\emph{Boundary Clock} -- In the boundary clock solution, the 5G radio access network (RAN) has direct access to the TSN grandmaster clock. The 5G RAN provides timing information to user equipments (UEs) via its own signaling and procedures. Based on the timing information, the UE synchronizes TSN devices. 

\emph{Transparent Clock} -- In the transparent clock solution, time synchronization is achieved through exchange of PTP messages. Any intermediate 5G or TSN entity between a TSN grandmaster and a TSN device updates PTP messages to correct for the time spent in the entity. 

\section{Over-the-Air Time Synchronization Analysis} \label{OTA}

\subsection{System Model}
\begin{figure}
    \centering
    \includegraphics[scale=0.35]{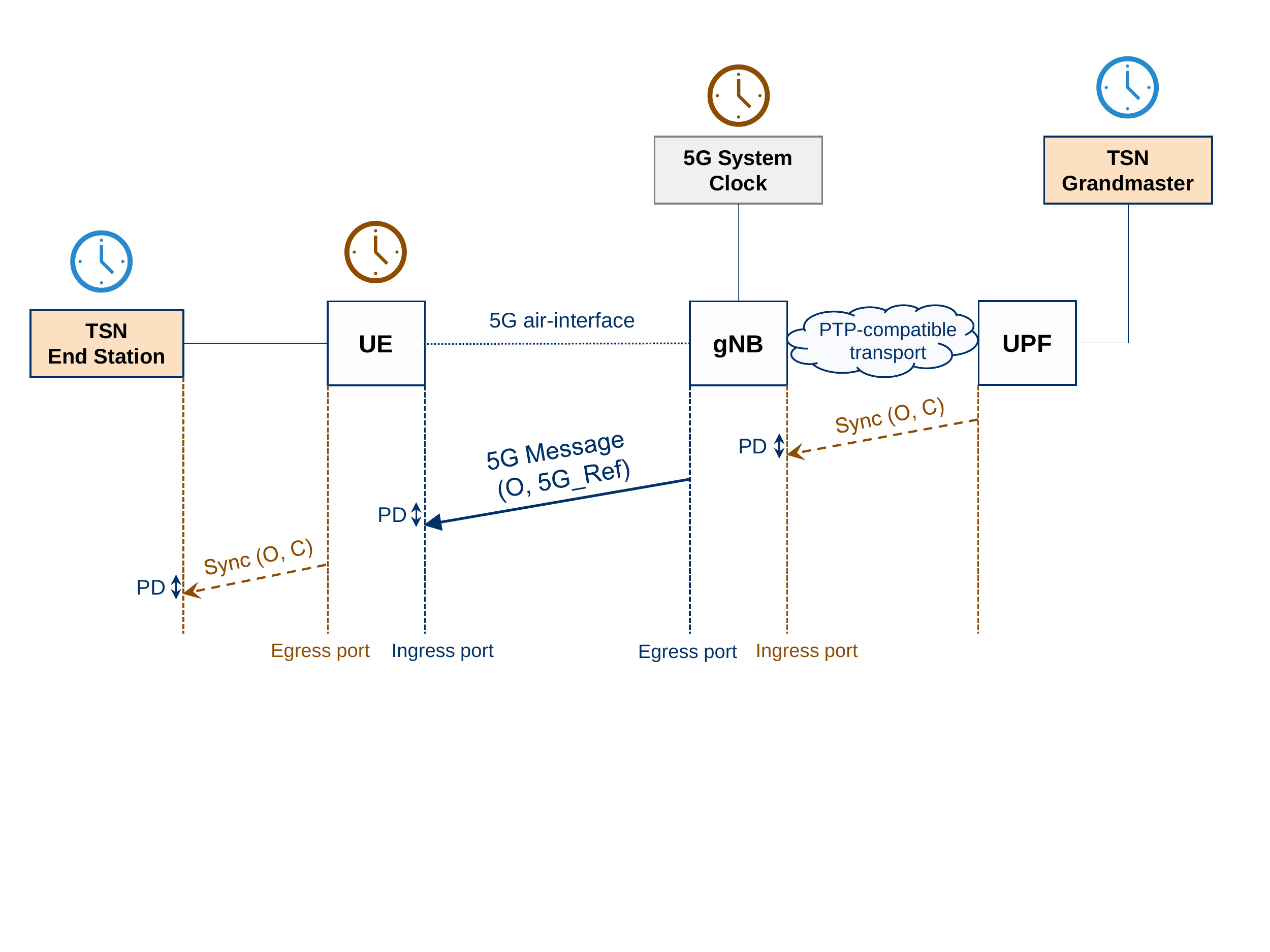}
    \caption{System model for over-the-air synchronization with boundary clock.}
    \label{sys_mod}
\end{figure}

We consider a boundary clock model for over-the-air time synchronization in integrated 5G and TSN systems as shown in Fig. \ref{sys_mod}. The 5G RAN has direct access to the TSN grandmaster time. This can be achieved either through direct connectivity with the TSN grandmaster clock or via the underlying PTP-capable transport network. We consider the latter scenario where the gNB is time synchronized to the TSN grandmaster time based on IEEE 802.1AS messages over the transport network providing interconnectivity with the user-plane function (UPF) located in the core network. The 5G RAN is time synchronized to the 5G system clock. The gNB provides TSN timing information over-the-air to the UE based on 5G broadcast/unicast signaling. The UE synchronizes the TSN end station based on IEEE 802.1AS messages. The 5G RAN indicates TSN timing information through a reference time (5G\(\_\)Ref) as per its fine frame structure. Note that timing information for multiple TSN domains can be provided through this approach. 

\begin{figure}
    \centering
    \includegraphics[scale=0.45]{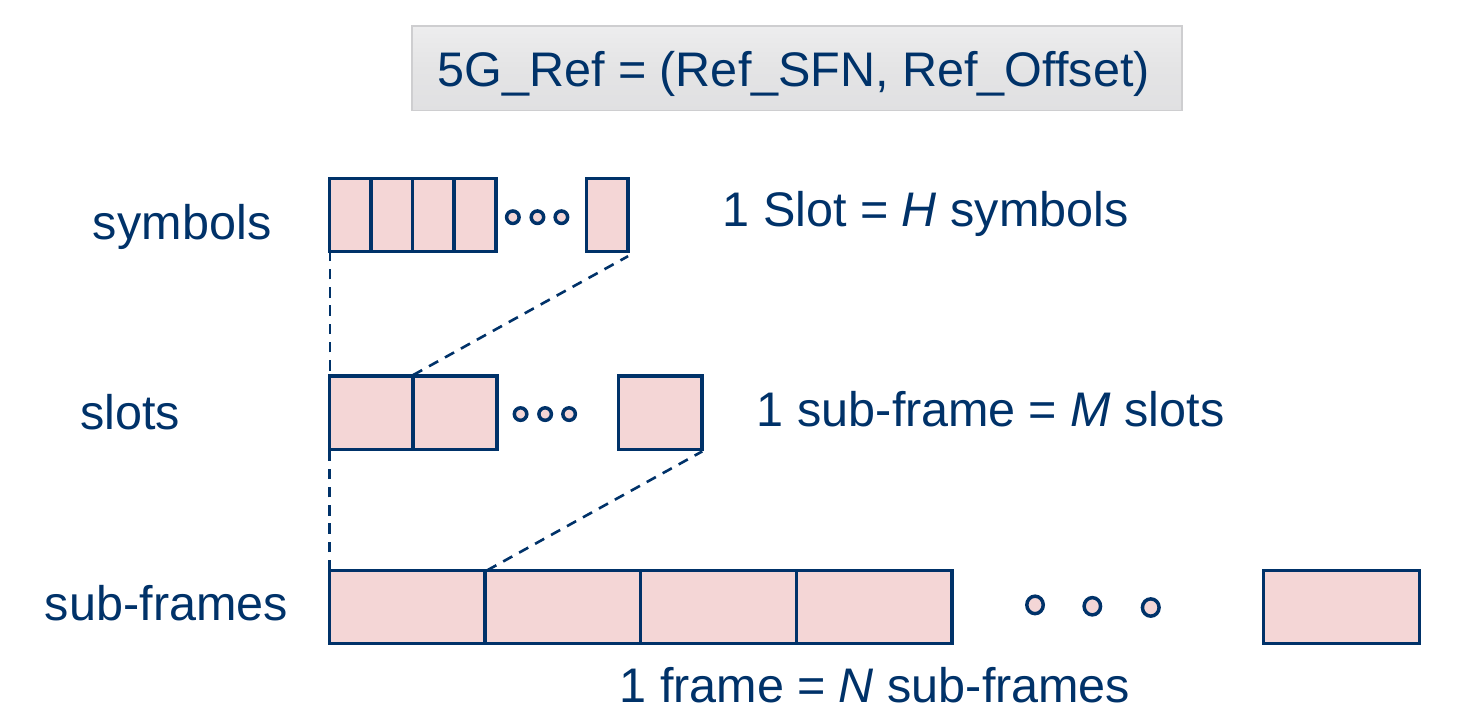}
    \caption{Illustration of reference time indication for TSN in 5G system.}
    \label{ref_time}
\end{figure}

After receiving an IEEE 802.1AS message, the gNB determines the time in 5G time base when the TSN grandmaster time the same as that indicated in the preciseOriginTimestamp (O). The gNB conveys 5G reference time along with the preciseOriginTimestamp (O) for accurate time synchronization of TSN bridges and/or end stations connected to the UE. The 5G air-interface is based on a time-slotted system of frames as depicted in Fig. \ref{ref_time}. Each frame consists of multiple sub-frames; each sub-frame comprises multiple slots and each slot further comprises a fixed number of OFDM symbols. As per 3GPP \cite{3gpp_23_734}, the reference time (5G\(\_\)Ref) consists of a reference sub-frame number (Ref\(\_\)SFN) and a reference offset (Ref\(\_\)Offset) within Ref\(\_\)SFN. Further, in accordance with ongoing activities within 3GPP, we assume that the gNB provides TSN timing information through system information blocks (SIBs). A potential candidate is the SIB16 \cite{3GPP_36_331} which contains a common time reference (GPS or UTC); however, its granularity needs to be enhanced for TSN synchronization. The SIBs are carried over the physical downlink shared channel (PDSCH). Resource allocation information for PDSCH transmissions is contained in the physical downlink control channel (PDCCH).

\subsection{Analytic Framework}
Let \(T_{\text{gNB}}\) denote the reference time (i.e., equivalent of the TSN grandmaster time) at the gNB. \(T_{\text{gNB}}\) is transmitted in a 5G message (e.g., in SIB) to the UE for time synchronization. The minimal synchronization frequency is \(f_\text{sync}\), which denotes time of synchronization with the granularity \(1/f_\text{sync}\) as \(t\). Let \(T_{\text{UE}}\) denote the local time at the UE, which is given by
\begin{align}\label{eq1}
T^t_{\text{UE}} = \left\{
\begin{aligned}
    & T^{t-1}_{\text{UE}} + \frac{1}{f_\text{sync}} +  \theta, & I^t_\text{sync} = 0, \\
    & T^t_{\text{gNB}} + \phi_{\text{PD}}+ \phi_\text{TAE} + \phi_\text{RTGE} , & I^t_\text{sync} = 1,  \\
\end{aligned} \right.
\end{align}
where \(\theta\) is the drift of the UE clock (compared to the gNB clock) per synchronization granularity, \( \phi_{\text{PD}}\) is the compensation of propagation delay, \( \phi_{\text{TAE}}\) is the compensation of time alignment error, and \( \phi_{\text{RTGE}}\) is the compensation of reference time granularity error. In \eqref{eq2}, $I^t_\text{sync}$ is a synchronization indicator at the current time $t$, where an updated reference time \(T_{\text{gNB}}\) is received at the UE if $I^t_\text{sync} = 1$, otherwise 0.

The time difference between clock of reference and the UE at time \(t\) is given by
\begin{align}\label{eq2}
X^t_{\text{TD}} = \left\{
\begin{aligned}
    & T^t_{\text{UE}} - T^t_{\text{gNB}}, & I^t_\text{sync} = 0, \\
    & T^t_{\text{UE}} - T^t_{\text{gNB}} + \text{PD}_{\text{gNB-to-UE}} + X_{\text{error}}, & I^t_\text{sync} = 1,  \\
\end{aligned} \right.
\end{align}
where \(\text{PD}_{\text{gNB-to-UE}}\) is the exact propagation delay from the gNB to the UE, and \(X_{\text{error}}\) is an error factor denoting cumulative errors associated with delivery of reference time, which is represented as
\begin{align}\label{eq2-1}
X_{\text{error}} = X_\text{TAE} + X_\text{RTGE} + X_\text{ToA}.
\end{align}
In \eqref{eq2-1}, \( X_{\text{TAE}}\) is the time alignment error, \( X_{\text{RTGE}}\) is the reference time granularity error, and \( X_{\text{ToA}}\) is the time of arrival error. All errors will be detailed in the next subsection.


\subsection{Path Delay Estimation and Errors} 
Estimation of path (propagation) delay from the gNB to the UE is an important factor for accurate time synchronization. However, in case of reference time indication via over-the-air messages, there is no explicit propagation delay measurement procedure as in IEEE 802.1AS. 
The path delay from the gNB to the UE can be estimated in two different ways as shown in \eqref{path_delay}. 
\begin{equation}
\label{path_delay}
\text{PD}_{\text{gNB-to-UE}}=
\begin{cases}
\mathcal{R}/\mathcal{C} \quad \text{without TA-based estimation} \\
\text{TA}/\text{2} \quad \text{with TA-based estimation}
\end{cases}
\end{equation}

The first approach to estimate path delay is based on the cell radius \(\mathcal{R}\) and the speed of light \(\mathcal{C}\). However, cell radius information is not available at the UE. Moreover, estimating cell radius purely based on received power is not entirely accurate. The second approach is to estimate path delay from the timing advance (TA) value \cite{3gpp38133} which is assigned to the UE by the gNB. Note that TA command is a MAC control element, which is responsible to align the time of sub-carriers in the uplink. Delivery of TA value to the UE is an implementation-specific issue, i.e., the gNB decides when to deliver TA information to the UE. For the sake of our analysis, we assume that the gNB is able to successfully deliver the TA command to the TSN UEs which have high accuracy synchronization requirements. 

The TA command for a connected UE\footnote{Connected UE refers to the UE was registered in the gNB. For an unconnected UE, random access procedure needs to be proceeded to build connection, and synchronization is initiated by transmitting an TA command via random access response with an TA index from $0,1,2,...,3846$.} contains one byte data, whereas the first two bits are set to 0, and the remaining 6 bits carries TA information representing a value $\Phi_\text{TA}$ with range from 0 to 63. The value of TA depends on the distance between the UE and the gNB (i.e., propagation delay) that is estimated by gNB. The compensation time $\phi_\text{PD}$ is obtained by 
\begin{align}\label{eq3}
\phi_\text{PD} = (\Phi_\text{TA} - 31) (N_\text{TA} + N_\text{TAfo}) T_{c},
\end{align}
In \eqref{eq3}, $T_{c}$ is the basic time unit for the 5G NR system, which is obtained by 
\begin{align}\label{eq4}
T_{c} = \frac{1}{\Delta F_\text{max} N_f},
\end{align}
where $\Delta F_\text{max}$ is sub-carrier spacing (SCS) and $N_f$ is Fast Fourier transform (FFT) size. In \eqref{eq3}, $ N_\text{TAfo}$ is a constant value denoting the frequency offset defined by 3GPP standards \cite{3gpp38133}. And, $N_\text{TA}$ is the number of time unit for 16 samples when time alignment, which is obtained by
\begin{align}\label{eq5}
N_\text{TA} = 16\times64\times2^\mu,
\end{align}
where $\mu = 0, 1, 2, \cdots$ denotes the index of NR numerology, and the SCS is $2^\mu \times 15$ kHz. 

As elaborated in \eqref{eq3}, compensation time of the propagation delay $\phi_\text{PD}$ is obtained by TA. However, it is limited by its granularity which is time slot $T_\text{gran}=((N_\text{TA} + N_\text{TAfo}) T_{c})/2$. Hence a maximum synchronization error of $\pm T_\text{gran}/2$ is introduced to the propagation delay compensation. In addition, the random errors in the original time of arrival value may result in a wrong TA  selection which will have a greater negative impact to the estimation as discussed in section \ref{toaee}.

\subsection{Time Alignment Error}
The time alignment error (TAE) can be referred to as the the frame timing accuracy of the gNB which is typically defined as a requirement. This requirement is defined as the frames of 5G-NR signals at the gNB transmit antenna connectors may experience timing differences with respect to each other. 3GPP has defined different requirements of TAE for different scenarios which are summarized as follows \cite{3gpp_38_104}. The most stringent requirement is for MIMO and transmit diversity where the TAE must be within \(\pm\)65 ns.

\subsection{Reference Time Granularity Error}
The granularity of reference time directly affects its accuracy compared to the TSN grandmaster clock. On average, the granularity of 5G\(\_\)Ref contributes an error of \(\pm {G}_{R}/2\), where \({G}_{R}\) denotes the reference time granularity as determined by Ref\(\_\)SFN and Ref\(\_\)Offset parameters. 

\subsection{Time of Arrival Estimation Error}\label{toaee}

An important factor in time of arrival (ToA) estimation is the timing of the downlink frame at the UE. It includes the detection error of the downlink signal at the UE and its internal processing jitter. According to 3GPP \cite{3gpp38133}, such UE-related timing errors have different values in different scenarios, defined by
\begin{equation}
\label{UE_tim_error}
X_{\text{ToA}}=
\begin{cases}
\pm 12 \times 64 \times T_c, \qquad \text{15 kHz}, \\
\pm 10 \times 64 \times T_c, \qquad \text{30 kHz}, \\
\pm 7 \times 64 \times T_c, \ \quad  \ \ \ \ \text{60 kHz}, \\
\pm 3.5 \times 64 \times T_c, \qquad \text{120 kHz}.
\end{cases}
\end{equation}

However, since the ToA is perturbed by both the noise and multipath error in densely cluttered environments \cite{mahmood2019time}, a more practical analysis should model the ToA error by considering exact channel fading. As evaluated in \cite{gentile2012geolocation}, the ToA error $X_{\text{ToA}}$, when experiencing an line-of-sight (LoS) channel, can be modeled by a zero-mean Gaussian distribution, where the variance depends on the multi-path structure. However, modeling the non-LOS (NLoS) multi-path environment is more challenging since, when a direct path being buried by noise, a non-direct path that generally experiences longer propagation path might be detected, which introduces bias to the Gaussian ToA error model. 
In this paper, we consider the ToA error is modelled by a zero-mean Gaussian distribution in a LoS channel environment. The variance is denoted by $\sigma$ that is proportional to the length of 5G time unit and SCS, which is represented as
\begin{align}\label{eq6}
\sigma=\frac{(N_\text{TA} + N_\text{TAfo}) T_{c}}{\kappa},
\end{align}
In \eqref{eq6}, $\kappa$ is a constant factor related to channel environment.

\section{Performance Evaluation} \label{perf}
Numerical simulations are conducted, based on our analytic framework, for evaluating the performance of over-the-air time synchronization. We evaluate the results by considering 5G numerology with one or several SCS values ($\mu$), including 15, 30, 60 and 120 kHz. According to the numerology setting of 5G NR system \cite{3gpp38133}, the maximal SCS $\Delta F_\text{max}$ is 480 kHz and the FFT size $N_f$ is 4096, which consequently conducts to a time unit covering 0.509 ns. We set the clock drift of the UE clock (compared to the gNB clock), i.e., $\theta$, to 10 ppm  ($10$ ns per ms). 

\begin{figure}
    \begin{center}
    \begin{minipage}[h]{0.5\textwidth}
    \centering
         \includegraphics[width=0.85\textwidth]{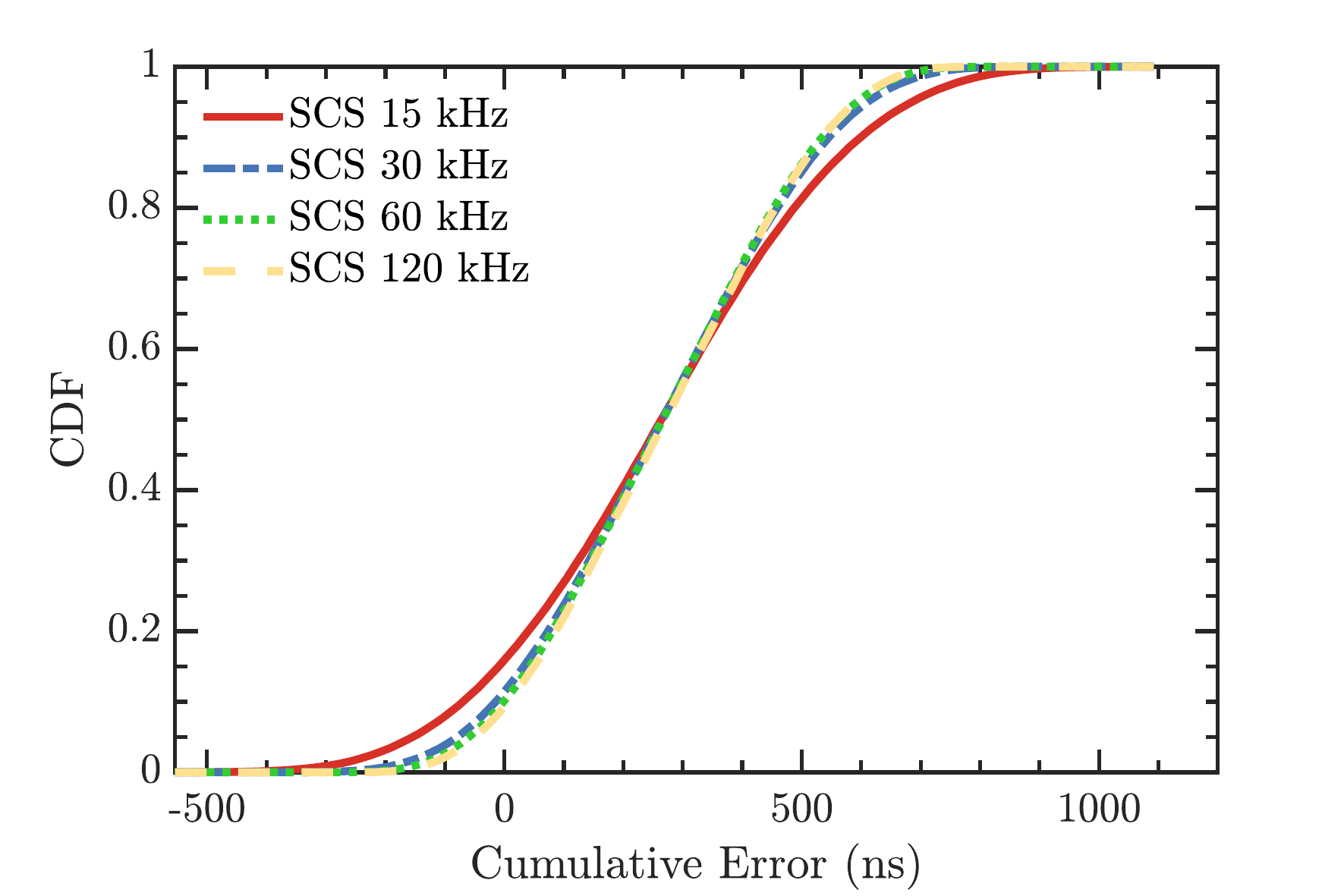}
        \caption{CDF of the cumulative time synchronization error.}
         \label{fig4}
        \end{minipage}
    \end{center}
\end{figure}

Fig. \ref{fig4} shows the cumulative distribution function (CDF) of the cumulative (overall) error in time synchronization for different values of SCS in the presence of path delay estimation, time alignment, reference time indication and ToA estimation errors. For this CDF, we assume that the reference time granularity is randomly distributed in [10 300] ns. Moreover, frequency of synchronization messages is fixed to 60 ms and ToA estimation is based on Gaussian distribution. The results indicate that a higher SCS leads to a lower error in time synchronization. This is mainly because of reduced slot duration at higher SCS. The results also indicate that the 99.9th-percentile of error is below 1 \(\mu\)s for SCS of 15 kHz, whereas for other SCS values, it is the 99.999th-percentile. 

\tablename~\ref{timeSyncError} summarizes the impact of path delay estimation and ToA estimation errors based on the CDF in Fig. \ref{fig4}. It captures the absolute mean error, the mean error and the maximum error for different values of SCS. The errors reduce significantly with higher SCS. From our evaluation of ToA estimation based on the Gaussian distribution, we observe that the mean error is not centred at zero. Therefore, we added an error correction field of  \(-\sigma/2\) as a complement the path delay compensation. The correction field has a positive impact on the error performance. The maximum error is reduced by approximately 50\%. The maximum error is now kept under $1\mu s$ even when $\kappa=1$ and SCS is 15kHz. Thus, our proposed path delay estimation is more tolerant to random errors arising from the multi-path wireless channels.

\begin{table}
 \caption{Path delay estimation and ToA estimation errors}
 \label{timeSyncError}
\begin{center}
\begin{tabular}{ |c|c|c|c|c| } 
 \hline
 Sub-carrier Spacing & 15kHz & 30kHz & 60kHz &120kHz \\ 
 \hline
 \multicolumn{5}{|c|}{$\kappa = 2$} \\
 \hline
 Absolute Mean Error (ns) & 192 & 96 & 48 & 24 \\
 Mean Error (ns) & 129 & 65 & 32 & 16 \\
 Maximum Error (ns) & 919 & 465 & 234 & 123 \\
 \hline
 \multicolumn{5}{|c|}{$\kappa= 1$} \\
 \hline
 Absolute Mean Error (ns) & 261 & 131 & 65 & 32 \\
 Mean Error (ns) & 129 & 65 & 32 & 16 \\
 Maximum Error (ns) & 1495 & 795 & 395 & 192 \\
 \hline
 \multicolumn{5}{|c|}{$\kappa = 2$ with error correction $-\sigma/2$} \\
 \hline
 Absolute Mean Error (ns) & 138 & 69 & 35 & 17 \\
 Mean Error (ns) & 1 & 0 & 0 & 0 \\
 Maximum Error (ns) & 557 & 270 & 138 & 67 \\
 \hline
 \multicolumn{5}{|c|}{$\kappa = 1$ with error correction $-\sigma/2$} \\
 \hline
 Absolute Mean Error (ns) & 162 & 81 & 40 & 20 \\
 Mean Error (ns) & 0 & 0 & 0 & 0 \\
 Maximum Error (ns) & 888 & 442 & 212 & 105 \\
 \hline
 \end{tabular}
\end{center}
\end{table}


\begin{figure}
    \begin{center}
    \begin{minipage}[h]{0.5\textwidth}
    \centering
         \includegraphics[width=0.85\textwidth]{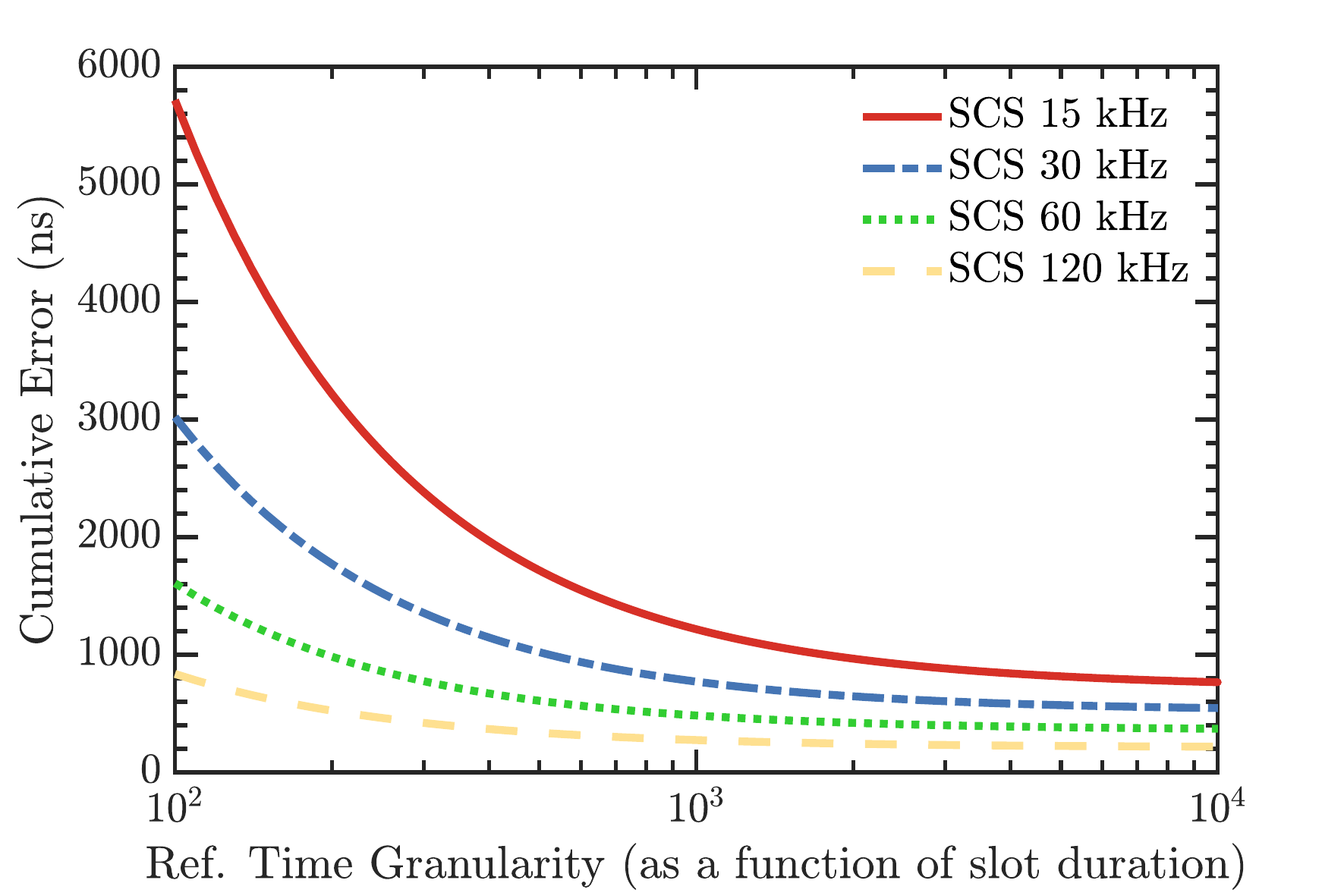}
        \caption{Cumulative error against granularity of reference time.}
         \label{fig5}
        \end{minipage}
    \end{center}
\end{figure}

Fig. \ref{fig5} plots cumulative error against reference time granularity for different SCS values. Note that the granularity of reference time is defined as a function of slot duration. For example, \(10^2\) indicates reference time granularity of 1/100th of slot duration. For this result, we have used 3GPP-defined values of UE timing errors, given by \eqref{UE_tim_error}, as ToA estimation under different values of SCS. The results indicate that reference time granularity must be on the order of tens of ns to ensure that the overall error remains below 1 \(\mu\)s in the presence of other errors associated with reference time indication.

\begin{figure}
    \begin{center}
    \begin{minipage}[htbp]{0.5\textwidth}
    \centering
        \includegraphics[width=0.85\textwidth]{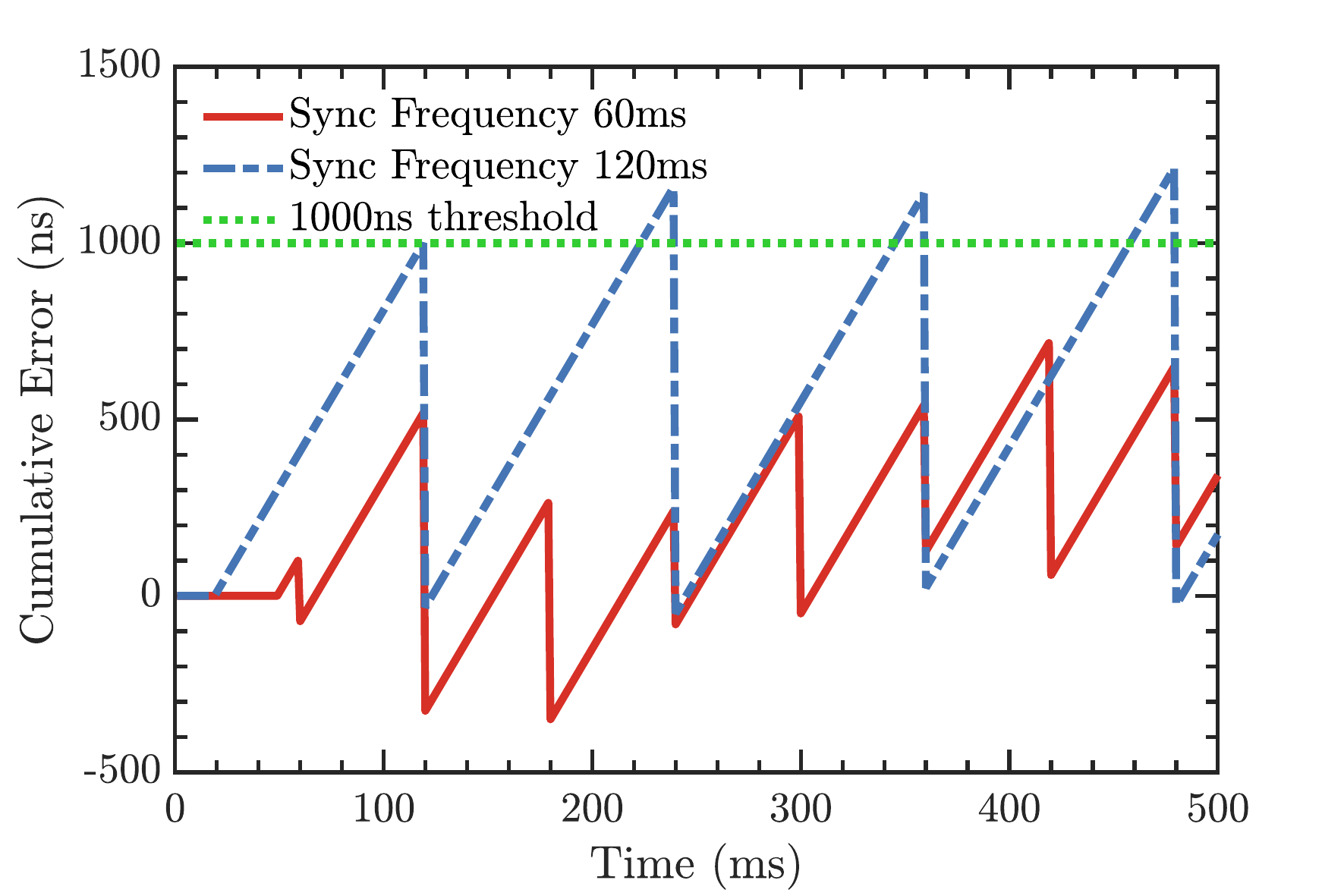}
        \caption{Time Difference between the clock of UE and gNB.}
                \label{SlaveClockDrift}
        \end{minipage}
    \end{center}
\end{figure}

Next, we investigate the impact of the frequency of timing messages from the gNB on synchronization performance. We start by illustrating the cumulative error between UE and gNB during the operation of synchronization. Fig.~\ref{SlaveClockDrift} plots the simultaneously cumulative error along time when synchronization messages are transmitted every 60 and 120 ms. Note that, to guarantee the reliability, the cumulative error generally needs to be less than 1 \(\mu\)s or 1000 ns \cite{3gpp22804}, which is showed by the green dotted line. As can be seen for each curve, the cumulative error continuously increases along time due to the positive clock drift, while the cumulative error is sharply reduced every 60 and 120 ms due to the periodic synchronization. We can observe that only the curve with 60 ms synchronization period can maintain the cumulative error that is always smaller than the 1000 ns threshold.

\begin{figure}
    \begin{center}
    \begin{minipage}[htbp]{0.5\textwidth}
    \centering
        \includegraphics[width=0.85\textwidth]{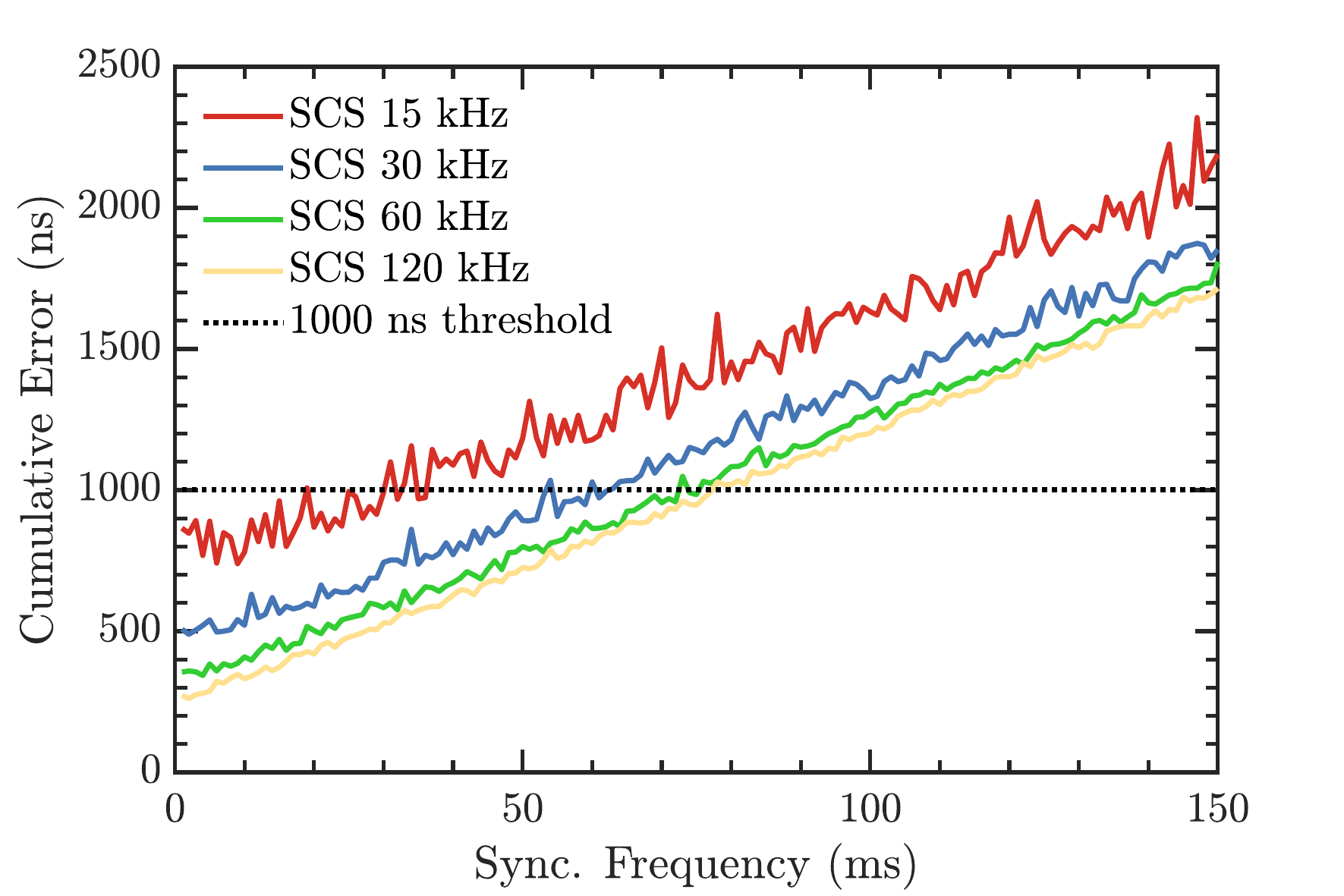}
                \label{SyncFreq}
        \end{minipage}
        \caption{Maximum time difference between the clock of UE and gNB versus  synchronization frequency.}
    \end{center}
\end{figure}

Fig.~\ref{SyncFreq} presents the maximum cumulative error between UE and gNB versus different synchronization period from 1 to 150 ms. We can observe the trend of maximal cumulative error increases along the synchronization period, since reducing synchronization frequency leads to larger accumulated clock drift between two consecutive synchronization chances. We further observe that the maximal cumulative error for each synchronization period increases with the decrease of SCS, which indicates that higher SCS is more robust to the cumulative error. One interesting result is that the gaps of maximal cumulative error between each two consecutive curves decrease along the order from SCS (15, 30) to (60, 120) kHz. This sheds light on the benefit of increasing SCS for keeping cumulative error of synchronization continuously decaying.

\begin{figure}
    \begin{center}
    \begin{minipage}[htbp]{0.45\textwidth}
    \centering
        \includegraphics[width=0.98\textwidth]{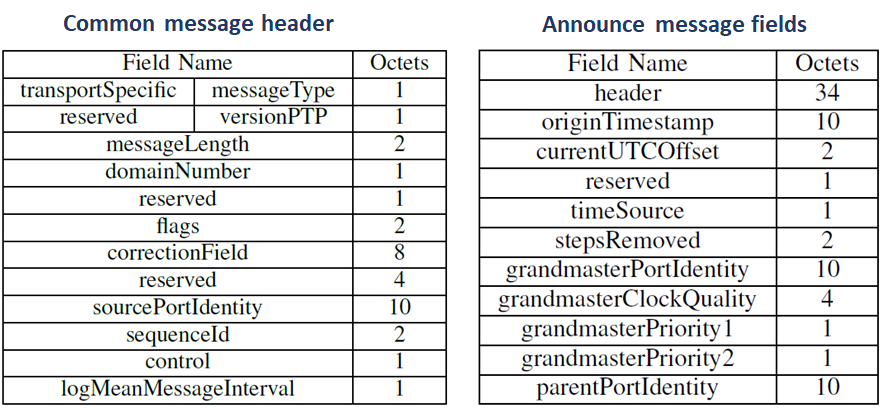}
        \caption{Message header and fields of the 802.1AS message.}
                \label{msg}
        \end{minipage}
    \end{center}
\end{figure}

Finally, we investigate the scalability aspects of over-the-air time synchronization as this approach is capable of supporting multiple TSN domains, each with its own TSN grandmaster clock. Fig. \ref{msg} illustrates the common message header and announce message fields of 802.1AS message. A 5G time synchronization message from the gNB containing TSN timestamp (originTimestamp) and the message header create a payload of 352 bits. Since the maximum size of a SIB message is 2976 bits, over-the-air time synchronization can support a maximum of 8 TSN domains.

\section{Concluding Remarks} \label{CR}
Seamless (tight) integration of 5G and TSN systems  is not possible without accurate time synchronization. We have investigated the performance of over-the-air time synchronization technique which is particularly important for the bridge model of 5G/TSN integration. We have developed an analytic framework that accounts for different errors associated with reference time indication. Our results reveal the impact of different errors on time synchronization performance and the conditions under which 1 \(\mu\)s or below synchronization requirement can be realized. In particular, path delay estimation and time of arrival estimation errors may have a significant impact on accuracy. The granularity of reference time must be on the order of 10 ns. Moreover, frequency of time synchronization plays an important role in keeping the clock drift within limits. Our future work will broadly focus on techniques for improving the accuracy and the efficiency of  over-the-air time synchronization.

\bibliographystyle{IEEEtran}
\bibliography{IEEEabrv,TSN_bib}

\end{document}